\documentclass[11pt,a4paper]{article}
\usepackage{jheppub_kim}

\usepackage{pdflscape}
\usepackage{amsmath}
\usepackage{amssymb}
\usepackage{dcolumn}
\usepackage{bm}
\usepackage{color}
\usepackage{epsfig}
\usepackage{amsfonts}
\usepackage{graphicx}
\usepackage{subfigure}
\usepackage{dcolumn}

\PassOptionsToPackage{linktocpage}{hyperref}
\def\case#1/#2{\textstyle\frac{#1}{#2}}
\newcommand{\beq}{\begin{equation}}
\newcommand{\eeq}{\end{equation}}
\newcommand{\bea}{\begin{eqnarray}}
\newcommand{\eea}{\end{eqnarray}}

\def\be{\begin{equation}}
\def\ee{\end{equation}}
\def\bea{\begin{eqnarray}}
\def\eea{\end{eqnarray}}

\begin{document}

\title{ Gravity's Rainbow: a bridge towards Ho\v{r}ava-Lifshitz gravity }

\author[a,b]{Remo Garattini}

\author[c,d]{Emmanuel N. Saridakis}

\affiliation[a]{Universit\`{a} degli Studi di Bergamo, Facolt\`{a} di Ingegneria, Viale
Marconi 5, 24044 Dalmine (Bergamo) Italy}

\affiliation[b]{I.N.F.N. - sezione di Milano, Milan, Italy}

\affiliation[c]{Institut d'Astrophysique de Paris, UMR 7095-CNRS,
Universit\'e Pierre \& Marie Curie, 98bis boulevard Arago, 75014 Paris,
France}

\affiliation[d]{Instituto de F\'{\i}sica, Pontificia
Universidad de Cat\'olica de Valpara\'{\i}so, Casilla 4950,
Valpara\'{\i}so, Chile}

\emailAdd{Remo.Garattini@unibg.it}

\emailAdd{Emmanuel$_-$Saridakis@baylor.edu}

\abstract{We investigate the connection between Gravity's Rainbow and Ho\v{r}ava-Lifshitz gravity,
since both theories incorporate a modification in the UltraViolet regime which improves their quantum behavior
at the cost of the  Lorentz invariance loss. In particular, extracting the Wheeler-De Witt equations of the two theories
in the case of Friedmann-Lema\^{\i}tre-Robertson-Walker and spherically symmetric geometries, we establish
a correspondence that bridges them. }


\maketitle

\newpage

\section{Introduction}

The idea that General Relativity (GR) is not the fundamental gravitational
theory and that needs to be modified or extended is quite old. On the one
hand, the idea of a small-scale, UltraViolet (UV) modification of GR arises
from the non-renormalizability of the theory and the difficulties towards its
quantization \cite{Stelle:1976gc}. In particular, since the usual loop-expansion 
procedure gives rise to UV-divergent Feynman diagrams, the
requirement for a UV-complete gravitational theory, which has GR as a
low-energy limit, becomes necessary. On the other hand, we know that the
large-scale, InfraRed (IR) modifications of GR might be the explanation of the
observed late-time universe acceleration (see \cite{Capozziello:2011et} and
references therein) and/or of the inflationary stage \cite{Nojiri:2003ft}. Due
to their significance, both directions led to a huge amount of research.

Concerning the modification of the UV behavior, it was realized that the
insertion of higher-order derivative terms in the Lagrangian establishes
renormalizability, since these terms modify the graviton propagator at high
energies \cite{Stelle:1976gc}. However, this leads to an obvious problem,
namely that the equations of motion involve higher-order time derivatives and
thus the application of the theory leads to ghosts. Nevertheless, based on the
observation that it is the higher spatial derivatives that improve
renormalizability while it is the higher time derivatives that lead to ghosts,
some years ago Ho\v{r}ava had the idea to construct a theory that allows for
the inclusion of higher spatial derivatives only. In order to achieve this,
and motivated by the Lifshitz theory of solid state physics \cite{Lifshitz},
he broke the ``democratic treating'' of space and time in the UV regime,
introducing an anisotropic, Lifshitz scaling between them
\cite{Horava2,Horava1,Horava3,Horava4}. Hence, higher spatial derivatives are
not accompanied by higher time ones (definitely this corresponds to Lorentz
violation), and thus in the UV the theory exhibits power-counting
renormalizability but still without ghosts. Finally, the theory presents
General Relativity as an IR fixed point, as required, where Lorentz invariance
is restored and space and time are handled on equal footing.

On the other hand, in \cite{MagSmo} the authors followed a different approach.
In particular, instead of modifying the action, they constructed an UV
modification of the metric itself, in a construction named Gravity's Rainbow
(GRw) \cite{MagSmo}. Hence, the deformed metric in principle exhibits a
different treatment between space and time in the UV, namely on scales near
the Planck scale, depending on the energy of the particle probing the
space-time, while at low energies one recovers the standard metric and General
Relativity is restored. Physically, one can think of it as a deformation of
the metric by the Planck-scale graviton. This deformation has been shown to to
cure divergences (at least to one loop) avoiding any
regularization/renormalization scheme \cite{GaMa,RJCAP}. Hence, due to this advantage,
a large amount of research has been devoted to Gravity's Rainbow 
 \cite{Galan:2004st,Hackett:2005mb,Ling:2005bp,Galan:2006by,Ling:2006ba,Weinfurtner:2008if,Li:2008gs,RemoGRw4,
 RemoGRw1,RemoGRw,RemoGRw2,RemoGRw3,RemoGRw8,RemoGRw7,RemoGRw6,Majumder:2013mza,Amelino-Camelia:2013wha,Awad:2013nxa,Barrow:2013gia,
 RemoGRw5,Ali:2014xqa,Ali:2014zea}.

In the present work we are interested in examining whether there is a
correspondence between Ho\v{r}ava-Lifshitz gravity and Gravity's Rainbow,
since both directions result in a modification of the equations in the UV
regime, while they both present GR as their low-energy limit. In particular,
since GR provides a natural scheme for quantization of the gravitational
field, namely the Wheeler-De Witt (WDW) equation \cite{DeWitt}, which is a
quantum version of the Hamiltonian constraint obtained from the
Arnowitt-Deser-Misner decomposition of space-time, we will impose that the WDW
equation must be satisfied by GRw and HL, respectively. We will examine this
correspondence on the Friedmann-Lema\^{\i}tre-Robertson-Walker (FLRW) metric
at the mini-superspace level, where the problem with the scalar graviton is
absent, as well as in spherically symmetric geometries.

The manuscript is organized as follows: in Section~\ref{p1} we review the
basic elements of Ho\v{r}ava-Lifshitz theory, while in Section~\ref{p2} we
extract the corresponding WDW equation in the case of FLRW space-time. In
Section~\ref{p3} we extract the WDW equation for Gravity's Rainbow in the case
of FLRW space-time. Then in Section \ref{p4} we establish the correspondence
between the two theories, while in Section \ref{p5} we obtain this relation
for spherically symmetric spacetimes. Finlly, we summarize our results in
Section \ref{p6}. Throughout this manuscript we use units in which
$\hbar=c=k=1$.

\section{Ho\v{r}ava-Lifshitz gravity}

\label{p1}

We start with a brief review of Ho\v{r}ava-Lifshitz gravity
\cite{Horava2,Horava1,Horava3,Horava4}. As we stated in the Introduction, the
central idea of the theory is the different treatment of space and time, which
allows us to introduce higher spatial derivatives without inserting also the
annoying higher time derivatives. Thus, a convenient framework to perform the
construction is the Arnowitt-Deser-Misner ($\mathcal{ADM}$) metric
decomposition, namely
\begin{align}
\label{metriciit}ds^{2} = - N^{2} dt^{2} + g_{ij} (dx^{i} + N^{i} dt ) (
dx^{j} + N^{j} dt ) .
\end{align}
 The dynamical variables are
the lapse $N$ and shift $N_{i}$ functions, and the spatial metric $g_{ij}$
(latin indices denote spatial coordinates). The coordinates scaling
transformations write as
\begin{equation}
t\rightarrow\ell^{3}t\ \text{\textrm{and}}\ x^{i}\rightarrow\ell x^{i},
\end{equation}
i.e. it is a Lifshitz scale invariance with a dynamical critical exponent
$z=3$.

The breaking of the four-dimensional diffeomorphism invariance allows for a
different treatment of the kinetic and potential terms for the metric in the
action, namely the kinetic term can be quadratic in time derivatives while the
potential term can have higher-order space derivatives. Thus, in general, the
action of Ho\v{r}ava-Lifshitz gravity is written as
\begin{equation}
S=\frac{1}{2\kappa}\int_{\Sigma\times I}dtd^{3}x\left(  \mathcal{L}%
_{K}-\mathcal{L}_{P}\right)  ,
\end{equation}
with $\kappa=M_{pl}^{-2}$ the Planck mass, where the kinetic term reads as
\begin{equation}
\mathcal{L}_{K}=N\sqrt{g}\left(  K^{ij}K_{ij}-\lambda K^{2}\right)  ,
\end{equation}
with $K_{ij}$ the extrinsic curvature defined as
\begin{equation}
K_{ij}=\frac{1}{2N}\left\{  -\dot{g}_{ij}+\nabla_{i}N_{j}+\nabla_{j}%
N_{i}\right\}  ,
\end{equation}
$K=K^{ij}g_{ij}$ its trace, and $g$ is the determinant of the spatial metric
$g_{ij}$. The constant $\lambda$ is a dimensionless running coupling, which
takes the value $\lambda=1$ in the IR limit. The potential part $\mathcal{L}%
_{P}$ can in principle contain many terms. However, one can make additional
assumptions in order to reduce the possible terms, thus resulting to various
versions of the theory. In the following we review the basic ones.

\subsection{Detailed balance version}

The assumption of \textquotedblleft detailed balance\textquotedblright%
\ \cite{Horava3} allows for the establishment of a quantum inheritance
principle \cite{Horava2}, that is the $(D+1)$-dimensional theory exhibits the
renormalization properties of the $D$-dimensional one. Physically, it
corresponds to the requirement that the potential term should arise from a
superpotential. This condition reduces significantly the potential part of the
action, resulting to
\begin{align}
&  \mathcal{L}_{Pdb}=N\sqrt{g}\left\{  \frac{\kappa^{2}}{w^{4}}C_{ij}%
C^{ij}-\frac{2\kappa^{3/2}\mu}{w^{2}}\frac{\epsilon^{ijk}}{\sqrt{g}}%
R_{il}\nabla_{j}R_{k}^{l}+\frac{\mu^{2}}{\kappa}R_{ij}R^{ij}\right.
\nonumber\\
&  \ \ \ \ \ \ \ \ \ \ \ \ \ \ \ \ \ \ \ \,\left.  - \frac{\mu^{2}}%
{1-3\lambda}\left[  \frac{1-4\lambda}{4}R^{2}+\Lambda R-\frac{3\Lambda^{2}%
}{\kappa}\right]  \right\}  , \label{potdb}%
\end{align}
where $C^{ij}=\epsilon^{ikl}\nabla_{k}\left(  R_{\ l}^{j}-\delta_{l\ }%
^{j}R/4\right)  /\sqrt{g}$ is the Cotton tensor (it is concomitant with the
metric and in three dimensions it is the analogue of the Weyl tensor), the
covariant derivatives are defined with respect to the spatial metric $g_{ij}$,
and $\epsilon^{ijk}$ is the totally antisymmetric unit tensor. Finally, apart
from the running coupling $\lambda$, we have three more constants, namely $w$,
$\mu$ and $\Lambda$. We mention that the detailed balance condition, apart
from reducing the possible terms in the potential part of the action, it
additionally correlates their coefficients, and thus the total number of
coefficients is smaller than the total number of terms.

\subsection{Projectable version}

Independently of the detailed balance condition one can impose the
\textquotedblleft projectability\textquotedblright\ condition, which is a weak
version of the invariance with respect to time reparametrizations, namely that
the lapse function is just a function of time, i.e. $N=N(t)$ \cite{Horava3}.
Such condition allows also for a significant reduction of terms in the
potential, since it eliminates the spatial derivatives of $N$. In this case,
and neglecting parity-violating terms, the potential part of the action
becomes \cite{Sotiriou:2009bx,Wang:2009yz}
\begin{align}
&  \mathcal{L}_{P}=N\sqrt{g}\left\{  g_{0}\kappa^{-1}+g_{1}R+\kappa\left(
g_{2}R^{2}+g_{3}R^{ij}R_{ij}\right)  \right. \nonumber\\
&  \ \ \ \ \ \ \,\left.  +\kappa^{2}\left(  g_{4}R^{3}+g_{5}RR^{ij}%
R_{ij}+g_{6}R_{j}^{i}R_{k}^{j}R_{i}^{k}+g_{7}R\nabla^{2}R+g_{8}\nabla
_{i}R_{jk}\nabla^{i}R^{jk}\right)  \right\}  , \label{lpnodb}%
\end{align}
where the couplings $g_{a}\left(  a=0\ldots8\right)  $ are all dimensionless
and running and moreover we can set $g_{1}=-1$. Finally, note that if apart
from the projectability condition one additionally imposes the detailed
balance condition, then he will again result in the potential term
(\ref{potdb}) but with $N=N(t)$.

\subsection{Non-projectable version}

In the general case where neither the detailed balance nor the projectability
conditions are imposed, one can have in the potential part of the action many
possible curvature invariants of $g_{ij}$, and moreover invariants including
also the vector $a_{i}=\partial_{i}\ln N$, which is now non-zero. In this case
the potential part of the action becomes \cite{Blas:2009qj}
\begin{equation}
\mathcal{L}_{Pnp}=N\sqrt{g}\left\{  -\xi R-\eta a_{i}a^{i}-\frac{1}{M_{A}^{2}%
}\mathcal{L}_{4}-\frac{1}{M_{B}^{2}}\mathcal{L}_{6}\right\}  ,
\end{equation}
where $a_{i}a^{i}$ is the lowest-order new term, of the same order with $R$,
and $\mathcal{L}_{4}$ and $\mathcal{L}_{6}$ respectively contain all possible
fourth and sixth order invariants that can be constructed by $a_{i}$ and
$g_{ij}$ and their combinations and contractions. Clearly, the above potential
term contains much more terms than the projectable or the detailed-balance
versions. Lastly, in order to recover GR in the IR limit, apart from the
running of $\lambda$ to 1, $\eta$ should run to zero too, while $\xi$ can be
set to 1.

We close this section by mentioning that in all versions of
Ho\v{r}ava-Lifshitz gravity, Lorentz invariance is violated due to both the
kinetic term (since $\lambda$ is in general not equal to 1) as well as to
terms in the potential. It is approximately and asymptotically restored in the
IR, where $\lambda$ runs to 1 and the potential terms will be significantly
suppressed. Thus, one can apply Ho\v{r}ava-Lifshitz gravity in order to
investigate its implications, which indeed are found to be rich and
interesting at both cosmological
\cite{Calcagni:2009ar,Kiritsis:2009sh,Brandenberger:2009yt,Nastase:2009nk,Mukohyama:2009zs,
Saridakis:2009bv,Mukohyama:2009mz,Wang:2009rw,Nojiri:2009th,Cai:2009in,Harko:2009rp,Yamamoto:2009tf,
Kobayashi:2009hh,Leon:2009rc,Wang:2009azb,Dutta:2009jn,RPRD,Cai:2010ud,Dutta:2010jh,
Kluson:2010aw,Jamil:2010di,Son:2010qh,Carloni:2010nx,Jamil:2010vr,Koutsoumbas:2010pt,Ali:2011sv,Koutsoumbas:2010yw,
Elizalde:2010ep,Saridakis:2011pk,Quevedo:2011fk,Abdujabbarov:2011uc,Izumi:2011eh,Saridakis:2012ui,Zhu:2011yu,Christodoulakis:2011np,
Briscese:2012rz,Zhu:2012zk,Maier:2013yh,Bellorin:2013zbp,Paul:2013vta,Alexandre:2013ura,Chattopadhyay:2014oba,Pasqua:2014ifa,Suresh:2014pra}
a well as black hole applications
\cite{Cai:2009pe,Kehagias:2009is,Cai:2009qs,Park:2009zra,Kiritsis:2009rx,Majhi:2009xh,Goldoni:2014oja}.

\section{The WDW equation in Ho\v{r}ava-Lifshitz gravity}

\label{p2}

In this section we examine the Wheeler-De Witt (WDW) equation in the framework
of Ho\v{r}ava-Lifshitz gravity. For convenience, and in order to simplify the
calculations, we focus on the projectable version of the theory, without the
detailed balanced condition, although an extension to the full,
non-projectable theory, is straightforward.

The WDW equation is a quantum version of the Hamiltonian constraint obtained
from the Arnowitt-Deser-Misner decomposition of space-time. Hence, let us
consider a simple mini-super-space model described by the FLRW line element%
\begin{equation}
ds^{2}=-N^{2}dt^{2}+a^{2}\left(  t\right)  d\Omega_{3}^{2}~, \label{FLRW}%
\end{equation}
describing a homogeneous, isotropic and closed universe. $d\Omega_{3}^{2}(k)$
is the metric on the spatial sections, which have constant curvature
$k=0,\pm1$, defined by
\begin{equation}
d\Omega_{3}^{2}=\gamma_{ij}dx^{i}dx^{j}~.
\end{equation}
Additionally, $N=N(t)$ is the lapse function and $a(t)$ denotes the scale
factor. In this background, the 3-dimensional Ricci curvature tensor and the
scalar curvature read
\begin{equation}
R_{ij}=\frac{2}{a^{2}\left(  t\right)  }\gamma_{ij}\qquad\mathrm{and}\qquad
R=\frac{6}{a^{2}\left(  t\right)  }~, \label{Curv}%
\end{equation}
respectively. With the help of Eq.$\left(  \ref{lpnodb}\right)  $, the
resulting Hamiltonian is computed by means of the usual Legendre
transformation, leading to
\begin{equation}
H=\int_{\Sigma}d^{3}x\mathcal{H}=\int_{\Sigma}d^{3}x\left[  \pi_{a}\dot
{a}-\mathcal{L}_{P}\right]  ,
\end{equation}
where $\pi_{a}$ is the canonical momentum. By inserting the FLRW background
into $\mathcal{L}_{P}$ one obtains
\begin{equation}
\mathcal{L}_{P}=N\sqrt{g}\left[  g_{0}\kappa^{-1}+g_{1}\frac{6}{a^{2}\left(
t\right)  }+\frac{12\kappa}{a^{4}\left(  t\right)  }\left(  3g_{2}%
+g_{3}\right)  +\frac{24\kappa^{2}}{a^{6}\left(  t\right)  }\left(
9g_{4}+3g_{5}+g_{6}\right)  \right]  . \label{LpFRLW}%
\end{equation}
The term $g_{0}\kappa^{-1}$ plays the role of a cosmological constant. In
order to make contact with the ordinary Einstein-Hilbert action in $3+1$
dimensions, we set without loss of generality%
\begin{align}
g_{0}\kappa^{-1}  &  \equiv2\Lambda\nonumber\\
g_{1}  &  \equiv-1. \label{coupling}%
\end{align}
Note that in the case where one desires to study the negative cosmological
constant, the identification will (trivially) be $g_{0}\kappa^{-1}\equiv-2\Lambda$ .

After having set $N=1$, the Legendre transformation leads to%
\begin{equation}
\mathcal{H}=\pi_{a}\dot{a}-\mathcal{L}_{K}+\mathcal{L}_{P},
\end{equation}
and the Hamiltonian constraint becomes \cite{RPRD}
\begin{align}
&  H=\int_{\Sigma}d^{3}x\mathcal{H}=-\frac{\kappa\pi_{a}^{2}}{12\pi
^{2}a\left(  3\lambda-1\right)  }+2\pi^{2}a^{3}\left(  t\right)  \left[
2\Lambda\kappa^{-1}-\frac{6\kappa^{-1}}{a^{2}\left(  t\right)  }+\frac
{12b}{a^{4}\left(  t\right)  }+\frac{24\kappa c}{a^{6}\left(  t\right)
}\right] \nonumber\\
&  \ \ \ \ \ \ \ \ \ \ \ \ \ \ \ \ \ \ =\pi_{a}^{2}+\frac{\left(
3\lambda-1\right)  }{\kappa^{2}}24\pi^{4}a^{4}\left(  t\right)  \left[
\frac{6}{a^{2}\left(  t\right)  }-\frac{12\kappa b}{a^{4}\left(  t\right)
}-\frac{24\kappa^{2}c}{a^{6}\left(  t\right)  }-2\Lambda\right]  =0,
\label{HHL}%
\end{align}
where%
\begin{align}
&  3g_{2}+g_{3}=b\nonumber\\
&  9g_{4}+3g_{5}+g_{6}=c. \label{bcgs}%
\end{align}
General Relativity is recovered when $b=c=0$, which does not necessarily means
that all the couplings are vanishing. Moreover, all the higher-curvature terms
are automatically suppressed, since the curvature becomes small
\cite{Sotiriou:2009bx}. Let us mention here that the scenario described by the
distorted potential Lagrangian $\left(  \ref{lpnodb}\right)  $, in the
specific case of FLRW geometry that we are interested in, could be considered
to arise equivalently in the framework of $f\left(  R\right)  $ gravity, with
$R$   the three-dimensional scalar curvature \cite{RJCAP}. Indeed, if ones
starts from the Lagrangian
\begin{equation}
\mathcal{L}_{fR}=N\sqrt{g}f\left(  R\right)  \label{Lpf(R)}%
\end{equation}
with
\begin{align}
f\left(  R\right)   &  =g_{0}\kappa^{-1}+g_{1}R-\frac{\kappa b}{3}R^{2}%
-\frac{\kappa^{2}c}{9}R^{3},\nonumber\\
&  =2\Lambda+R\left(  1-2\pi b\frac{R}{R_{0}}-4\pi^{2}c\frac{R^{2}}{R_{0}^{2}%
}\right)  , \label{f(R)}%
\end{align}
and $b$ and $c$ given by (\ref{bcgs}), and extract the corresponding field
equations in the case of FLRW geometry, he will obtain the same equations with
those extracted from $\mathcal{L}_{P}$ in $\left(  \ref{lpnodb}\right)  $.
Lastly, note that we have used the definitions $\left(  \ref{coupling}\right)
$, while we have furthermore set $R_{0}\equiv6/G=6/l_{p}^{2}$.

\section{The WDW equation in Gravity's Rainbow}

\label{p3}

In this section we review briefly the gravity's rainbow (GRw) \cite{MagSmo},
focusing on the Hamiltonian analysis and the WDW equation. In this
formulation, the space-time geometry is described by the deformed metric
\begin{equation}
ds^{2}=-\frac{N^{2}\left(  t\right)  }{g_{1}^{2}\left(  E/E_{\mathrm{Pl}%
}\right)  }dt^{2}+\frac{a^{2}\left(  t\right)  }{g_{2}^{2}\left(
E/E_{\mathrm{Pl}}\right)  }d\Omega_{3}^{2}~, \label{FLRWMod}%
\end{equation}
where $g_{1}(E/E_{\mathrm{Pl}})$ and $g_{2}(E/E_{\mathrm{Pl}})$ are functions
of energy, which incorporate the deformation of the metric. Concerning the
low-energy limit it is required to consider
\begin{equation}
\lim_{E/E_{\mathrm{Pl}}\rightarrow0}g_{1}\left(  E/E_{\mathrm{Pl}}\right)
=1\qquad\mathrm{and}\qquad\lim_{E/E_{\mathrm{Pl}}\rightarrow0}g_{2}\left(
E/E_{\mathrm{Pl}}\right)  =1,
\end{equation}
and thus to recover the usual FLRW geometry. Hence, $E$ quantifies the energy
scale at which quantum gravity effects become apparent. For instance, one of
these effects would be that the graviton distorts the background metric as we
approach the Planck scale.

As it has been extensively shown in the literature
\cite{GaMa,RJCAP,Galan:2004st,Hackett:2005mb,Ling:2005bp,Galan:2006by,Ling:2006ba,Weinfurtner:2008if,Li:2008gs,RemoGRw4,
 RemoGRw1,RemoGRw,RemoGRw2,RemoGRw3,RemoGRw8,RemoGRw7,RemoGRw6,Majumder:2013mza,Amelino-Camelia:2013wha,Awad:2013nxa,Barrow:2013gia,
 RemoGRw5,Ali:2014xqa,Ali:2014zea}, GRw can
be used to cure or alleviate the usual GR divergences, at least to one loop,
avoiding any regularization and renormalization schemes. If we allow the
energy $E$ to evolve depending on $t$, one finds that the extrinsic curvature
of the metric (\ref{FLRWMod}) reads
\begin{align}
K_{ij}  &  =-\frac{g_{1}\left(  E\left(  a\left(  t\right)  \right)
/E_{P}\right)  }{2N}\frac{d}{dt}\left[  \frac{g_{ij}}{g_{2}^{2}\left(
E\left(  a\left(  t\right)  \right)  /E_{P}\right)  }\right] \nonumber\\
&  =\frac{g_{1}\left(  E\left(  a\left(  t\right)  \right)  /E_{P}\right)
}{g_{2}^{2}\left(  E\left(  a\left(  t\right)  \right)  /E_{P}\right)
}\left[  \tilde{K}_{ij}+\tilde{g}_{ij}\frac{A\left(  t\right)  }{N\left(
t\right)  }\dot{a}\left(  t\right)  \right]  ~, \label{ExrCurvGrw}%
\end{align}
where%
\begin{equation}
A\left(  t\right)  =\frac{1}{g_{2}\left(  E\left(  a\left(  t\right)  \right)
/E_{P}\right)  E_{P}}\frac{d}{dE}\Big[g_{2}\left(  E\left(  a\left(  t\right)
\right)  /E_{P}\right)  \Big]\frac{dE}{da},
\end{equation}
and with dots denoting differentiation with respect to time. In the above
expressions   the tildes indicate the quantities computed in absence
of the rainbow's functions. 

The next step is to find the corresponding canonical momentum. After a short
calculation, presented in Appendix \ref{App}, the canonical momentum writes
as
\begin{equation}
\pi_{a}=\frac{\delta S_{K}}{\delta\dot{a}}=\frac{g_{1}^{2} \left(  E\left(
a\left(  t\right)  \right)  /E_{P}\right)  }{g_{2}^{3}\left(  E\left(
a\left(  t\right)  \right)  /E_{P}\right)  }\,f\!\left(  A\left(  t\right)
,a\right)  \tilde{\pi}_{a},
\end{equation}
where%
\begin{equation}
f\!\left(  A\left(  t\right)  ,a\right)  =\left[  1-2a\left(  t\right)
A\left(  t\right)  +A^{2}\left(  t\right)  a\left(  t\right)  ^{2}\right]  ,
\end{equation}
and where
\begin{equation}
\tilde{\pi}_{a}=\frac{6\pi^{2}}{\kappa}\frac{\left(  1-3\lambda\right)
}{N\left(  t\right)  }\dot{a}a.
\end{equation}

Finally, we can now assemble the Hamiltonian density, which is defined as
\begin{equation}
\mathcal{H}=\pi_{a}\dot{a}-\mathcal{L}_{K}+\mathcal{L}_{P},
\end{equation}
 where  $\mathcal{L}_{P}$  is the potential term whose form
is  
\begin{equation}
\mathcal{L}_{P}=\frac{N\left(  t\right)  \sqrt{\tilde{g}}}{16\pi Gg_{2}\left(
E\left(  a\left(  t\right)  \right)  /E_{P}\right)  }\!{}\!\left[  \tilde
{R}-\frac{2\Lambda}{g_{2}^{2}\left(  E\left(  a\left(  t\right)  \right)
/E_{P}\right)  }\right]  ~. \label{Acca}%
\end{equation}
Concerning the kinetic term we have
\begin{align}
&  \mathcal{H}_{K}=\pi_{a}\dot{a}-\mathcal{L}_{K}=\frac{\kappa N\left(
t\right)  }{12\pi^{2}a}\left[  \frac{g_{2}^{3}\left(  E\left(  a\left(
t\right)  \right)  /E_{P}\right)  }{g_{1}^{2}\left(  E\left(  a\left(
t\right)  \right)  /E_{P}\right)  }\right]  \frac{\pi_{a}^{2}}{\left(
1-3\lambda\right)  f\!\left(  A\left(  t\right)  ,a\right)  }\nonumber\\
&  \ \ \ \ \ \,=\left[  \frac{\kappa N\left(  t\right)  }{12\pi^{2}a}\right]
\left[  \frac{\tilde{\pi}_{a}^{2}}{\left(  1-3\lambda\right)  }\right]
\left[  \frac{g_{1}^{2}\left(  E\left(  a\left(  t\right)  \right)
/E_{P}\right)  }{g_{2}^{3}\left(  E\left(  a\left(  t\right)  \right)
/E_{P}\right)  }\right]  f\!\left(  A\left(  t\right)  ,a\right) ,
\end{align}
thus  the classical Hamiltonian constraint reduces to
\begin{align}
&  \mathcal{H}=\frac{\kappa}{12\pi^{2}a}\frac{\tilde{\pi}_{a}^{2}}{\left(
1-3\lambda\right)  }\frac{g_{1}^{2}\left(  E\left(  a\left(  t\right)
\right)  /E_{P}\right)  }{g_{2}^{3}\left(  E\left(  a\left(  t\right)
\right)  /E_{P}\right)  }f\!\left(  A\left(  t\right)  ,a\right) \nonumber\\
&  \ \ \ \ \ \,\ -\frac{\pi^{2}a^{3}\left(  t\right)  }{\kappa g_{2}\left(
E\left(  a\left(  t\right)  \right)  /E_{P}\right)  }\left[  \frac{6}%
{a^{2}\left(  t\right)  }-\frac{2\Lambda}{g_{2}^{2}\left(  E\left(  a\left(
t\right)  \right)  /E_{P}\right)  }\right]  =0. \label{Hc}%
\end{align}
It is then straightforward to see that the Hamiltonian density reduces to
\begin{equation}
\mathcal{H}=\tilde{\pi}_{a}^{2}+\frac{12\left(  3\lambda-1\right)  \pi
^{4}a^{4}\left(  t\right)  }{\kappa^{2}g_{1}^{2}\left(  E\left(  a\left(
t\right)  \right)  /E_{P}\right)  \,f\!\left(  A\left(  t\right)  ,a\right)
}\!{}\!\left[  g_{2}^{2}\left(  E\left(  a\left(  t\right)  \right)
/E_{P}\right)  \frac{6}{a^{2}\left(  t\right)  }-2\Lambda\right]  =0,
\label{HGR}%
\end{equation}
where we have integrated out all   degrees of freedom apart from the scale factor.

\section{Correspondence of Gravity's Rainbow with Ho\v{r}ava-Lifshitz gravity}

\label{p4}

In the previous sections we have extracted the WDW equation in the cases of   Ho\v{r}ava-Lifshitz gravity
and 
Gravity's Rainbow, for a FLRW background, that
is expressions (\ref{HHL}) and (\ref{HGR}) respectively. Hence, observing
their forms we deduce that it is possible to create a formal correspondence
between the two formulations provided that
\begin{equation}
g_{1}^{2}\left(  E\left(  a\left(  t\right)  \right)  /E_{P}\right)
f\!\left(  A\left(  t\right)  ,a\right)  =1
\end{equation}
and%
\begin{equation}
g_{2}^{2}\left(  E\left(  a\left(  t\right)  \right)  /E_{P}\right)  \frac
{6}{a^{2}\left(  t\right)  }=\frac{6}{a^{2}\left(  t\right)  } \left[
1-\frac{2\kappa b}{a^{2}\left(  t\right)  }-\frac{4\kappa^{2}c}{a^{4}\left(
t\right)  }\right]  .
\end{equation}
Since we preserve the freedom to fix $g_{2}\left(  E\left(  a\left(  t\right)
\right)  /E_{P}\right)  $, we impose that
\begin{align}
g_{2}^{2}\left(  E\left(  a\left(  t\right)  \right)  /E_{P}\right)   &
=1-\frac{2b\kappa}{a^{2}\left(  t\right)  }-\frac{4\kappa^{2}c}{a^{4}\left(
t\right)  }\nonumber\\
&  =1-\frac{16bR}{R_{0}}-\frac{256cR^{2}}{R_{0}^{2}}, \label{ide}%
\end{align}
where $R_{0}$ has been defined in  $\left(  \ref{f(R)}\right)  $ as
$R_{0}\equiv6/G=6/l_{p}^{2}$. Although at first site identification
(\ref{ide}) seems to be imposed \textit{ad hoc}, it can be supported by
invoking the dispersion relation of a massless graviton which, as we show in
Appendix \ref{Appen}, for a FLRW background acquires the form
\begin{equation}
E^{2}=\frac{k^{2}}{a^{2}\left(  t\right)  },
\end{equation}
with $k$ the constant dimensionless radial wavenumber, and thus in the present
case of Gravity's Rainbow it is modified to
\begin{equation}
\frac{E^{2}}{g_{2}^{2}\left(  E\left(  a\left(  t\right)  \right)
/E_{P}\right)  }=\frac{k^{2}}{a^{2}\left(  t\right)  }. \label{grav}%
\end{equation}
Since the dispersion relation $\left(  \ref{grav}\right)  $ is valid at the
Planck scale too, we can write
\begin{equation}
\frac{E^{2}}{g_{2}^{2}\left(  E\left(  a\left(  t\right)  \right)
/E_{P}\right)  }\rightarrow\frac{E_{P}^{2}}{g_{2}^{2}\left(  E_{P}%
/E_{P}\right)  }=E_{P}^{2}=\frac{k^{2}}{a_{P}^{2}}.
\end{equation}
Hence, Eq.$\left(  \ref{ide}\right)  $ becomes
\begin{align}
g_{2}^{2}\left(  E\left(  a\left(  t\right)  \right)  /E_{P}\right)   &
=1-\frac{16b\pi R}{R_{0}}-\frac{256c\pi^{2}R^{2}}{R_{0}^{2}}\nonumber\\
&  =1-c_{1}\frac{E^{2}\left(  a\left(  t\right)  \right)  }{E_{P}^{2}}%
-c_{2}\frac{E^{4}\left(  a\left(  t\right)  \right)  }{E_{P}^{4}}. \label{g2E}%
\end{align}
Therefore we deduce that
\begin{equation}
E^{2}=R/6k^{2}%
\end{equation}
with
\begin{equation}
E_{P}^{2}=G^{-1},\qquad c_{1}=16b\pi\text{\qquad\textrm{and\qquad}}%
c_{2}=256c\pi^{2}.
\end{equation}

We mention here that the fact that a relation between the energy of a particle
and the scalar curvature can come into play directly in the metric, is not a
novelty. Indeed in \cite{Olmo:2011sw} the scalar curvature enters into the
metric via the trace of the Einstein's field equations connecting the
energy-momentum tensor with the $4D$ scalar curvature. Moreover, note that the
energy-momentum tensor has dimensions of energy density. Thus, and in order to
take the comparison on a general ground, one can assume that $g_{2}\left(
E\left(  a\left(  t\right)  \right)  /E_{P}\right)  $ can be represented by a
formal expansion in powers of $E/E_{P}$, identifying the coefficients order by
order. However, since in the present work we are comparing Gravity's Rainbow
with the Ho\v{r}ava-Lifshitz gravity with $z=3$, the formal Taylor expansion
is truncated at the second order.

\section{Correspondence in spherically symmetric backgrounds}

\label{p5}

The discussion on the WDW equations in Gravity's Rainbow and
Ho\v{r}ava-Lifshitz gravity of the previous section was presented in
homogeneous and isotropic backgrounds, namely on the FLRW metric. One could
wonder whether these results are an artifact of the space-time symmetries and
not of the features of the two theories. Thus, in the present section we
repeat the above analysis in the case of spherically symmetric backgrounds. In
particular we consider metrics of the class
\begin{equation}
ds^{2}=-N^{2}\left(  r\right)  \,dt^{2}+\frac{dr^{2}}{1-b(r)/r}+r^{2}%
\,(d\theta^{2}+\sin^{2}{\theta}\,d\phi^{2})\,, \label{ssm}%
\end{equation}
where $N(r)$ and $b(r)$ are arbitrary functions of the radial coordinate $r$,
denoted as the lapse function and the form function respectively. In this
case, the energies now depend on the shape function $b\left(  r\right)  $ and
the radial coordinate $r$, namely
\begin{align}
&  g_{1}\left(  E/E_{P}\right)  \equiv g_{1}\left(  E\left(  b\left(
r\right)  \right)  /E_{P}\right) \nonumber\\
&  g_{2}\left(  E/E_{P}\right)  \equiv g_{2}\left(  E\left(  b\left(
r\right)  \right)  /E_{P}\right)  .
\end{align}
Hence, the metric modification appears on a scalar curvature $R$ given by
\begin{equation}
R=g^{ij}R_{ij}=\frac{2b^{\prime}\left(  r\right)  }{r^{2}},
\end{equation}
 where the prime denotes derivative with respect to $r$,  and
  we have used the mixed Ricci tensor $R_{j\text{ }}^{a}$ with
components 
\begin{equation}
R_{j\text{ }}^{a}=\left\{  \frac{b^{\prime}\left(  r\right)  }{r^{2}}%
-\frac{b\left(  r\right)  }{r^{3}},\frac{b^{\prime}\left(  r\right)  }{2r^{2}%
}+\frac{b\left(  r\right)  }{2r^{3}},\frac{b^{\prime}\left(  r\right)
}{2r^{2}}+\frac{b\left(  r\right)  }{2r^{3}}\right\}  . \label{Ricci}%
\end{equation}
When GRw switches on, the line element $\left(  \ref{ssm}\right)  $ becomes
\begin{align}
&  ds^{2}=-\frac{N^{2}\left(  r\right)  }{g_{1}^{2}\left(  E\left(  b\left(
r\right)  \right)  /E_{P}\right)  }\,dt^{2}+\frac{dr^{2}}{g_{2}^{2}\left(
E\left(  b\left(  r\right)  \right)  /E_{P}\right)  \left(  1-b(r)/r\right)
}\nonumber\\
&  \ \ \ \ \ \ \ \ +\frac{r^{2}}{g_{1}^{2}\left(  E\left(  b\left(  r\right)
\right)  /E_{P}\right)  }\,(d\theta^{2}+\sin^{2}{\theta}\,d\phi^{2})\,,
\end{align}
and the scalar curvature transforms as 
\begin{eqnarray}
&  \!\!\!\!\!\!\!\!\!R\rightarrow\left[  1-{\frac{b\left(  r\right)  }{r}%
}\right]  \left\{  r^{4}\,g_{2}\left(  E\left(  b\left(  r\right)  \right)
\right)  {\tilde{R}}^{2}\left\{  \frac{d^{2}g_{2}\left(  E\left(  b\left(
r\right)  \right)  \right)  }{dE^{2}}\left[  \frac{dE\left(  b\left(
r\right)  \right)  }{db}\right]  ^{2}+\frac{dg_{2}\left(  E\left(  b\left(
r\right)  \right)  \right)  }{dE}\frac{d^{2}E\left(  b\left(  r\right)
\right)  }{db^{2}}\right\}  \right. 
\nonumber\\
&  \left.  \!\!\!\!\!\!\!\!\! \!\!\!\!\!\!\!\!\! \!\!\!\!\!\!\!\!\! \!\!\!\!\!\!\!\!\! \!\!\!\!\!\!\!\!\!\!\!\!\!  
-\frac{3}{2}r^{4}{\tilde
{R}}^{2}\left[  \frac{dE\left(  b\left(  r\right)  \right)  }{db}\right]
^{2}\left[  \frac{dg_{2}\left(  E\left(  b\left(  r\right)  \right)  \right)
}{dE}\right]  ^{2}\right. \nonumber\\
&  \left. \!\!\!\!\!\!\!\!\! \!\!\!\!\!\!\!\!\! \!\!\!\!\!\!\!\!\!\!   
+4\,g_{2}\left(  E\left(
b\left(  r\right)  \right)  \right)  \frac{dE\left(  b\left(  r\right)
\right)  }{db}\frac{dg_{2}\left(  E\left(  b\left(  r\right)  \right)
\right)  }{dE}\frac{d^{2}b\left(  r\right)  }{dr^{2}}\right\} \nonumber\\
&  \ \ \ \ \cdot g_{2}\left(  E\left(  b\left(  r\right)  \right)  \right)
\frac{dg_{2}\left(  E\left(  b\left(  r\right)  \right)  \right)  }{dE}%
\frac{dE\left(  b\left(  r\right)  \right)  }{db}\left[  -\frac{r^{3}}%
{2}\tilde{R}^{2}-3{b\left(  r\right)  \tilde{R}+4r\tilde{R}}\right]
+g_{2}^{2}{\left(  E\left(  b\left(  r\right)  \right)  \right)  \tilde{R},}
\label{ModR}%
\end{eqnarray}
where  the tildes indicate the quantities computed in absence
of the rainbow's functions.
Although this is not necessary, for
simplification we focus on the case where there is no explicit dependence of
$E$ on $b\left(  r\right)  $, that is we assume $dE\left(  b\left(  r\right)
\right)  /db=0$. In this case the scalar curvature simplifies to%
\begin{equation}
R\rightarrow{g_{2}^{2}\left(  E\left(  b\left(  r\right)  \right)
/E_{P}\right)  \tilde{R}.}%
\end{equation}
Since the extrinsic curvature $K_{ij}$ becomes
\begin{equation}
K_{ij}=-\frac{\dot{g}_{ij}}{2N}=\frac{g_{1}\left(  E\left(  b\left(  r\right)
\right)  /E_{P}\right)  }{g_{2}^{2}\left(  E\left(  b\left(  r\right)
\right)  /E_{P}\right)  }\tilde{K}_{ij}, \label{Kij}%
\end{equation}
even in this case the kinetic term does not contribute at the classical level
and the GRw distortion is completely encoded in the potential term. Hence, if
we assume the validity of Eq. $\left(  \ref{g2E}\right)  $ for the spherically
symmetric case too, we find
\begin{align}
g_{2}^{2}\left(  E\left(  a\left(  t\right)  \right)  /E_{P}\right)   &
=1+g_{2}\frac{E^{2}\left(  b\left(  r\right)  \right)  }{E_{P}^{2}}+g_{4}%
\frac{E^{4}\left(  b\left(  r\right)  \right)  }{E_{P}^{4}}\nonumber\\
&  =1+g_{2}\frac{R}{R_{0}}+g_{4}\frac{R^{2}}{R_{0}^{2}}.
\end{align}
Therefore, we conclude that one can establish a correspondence between GRw and
HL in the spherically symmetric geometries too. Although we have shown  
this correspondence in the case of scalar curvature, we expect it to hold in
the general case too, although such a feature is needed to be proven formally.

\section{Conclusions}

\label{p6}

In this work we were interested in exploring the connection between two
Lorentz-violating theories, namely Gravity's Rainbow and Ho\v{r}ava-Lifshitz
gravity. In Gravity's Rainbow, it is the metric that incorporates all the
distortion of the space time when one approaches the Planck scale, while in
Ho\v{r}ava-Lifshitz gravity, it is the potential part of the action (or the
Hamiltonian) that acquires higher-order curvature terms. Usually Gravity's
Rainbow switches on because a Planckian particle distorts the gravitational
metric tensor $g_{\mu\nu}$. However, since in the present application we have
neglected any matter fields, the only particle appearing is the graviton.
Since the graviton is the quantum particle associated with the quantum
fluctuations of the space time, we conclude that it is the gravitational field
itself that it is responsible for such a distortion. This is also enforced by
the dispersion relation relating the graviton energy and the scale factor,
namely the scalar curvature, in the case where an FLRW background is imposed,
or the graviton energy and the shape function  in the case where a spherically
symmetric background is imposed.

As we have shown, one can indeed establish a correspondence between the two
theories, through the examination of their Wheeler-De Witt equations. However,
although we have explicitly shown this in the case of two physically
interesting spacetimes, namely the FLRW and the spherically symmetric ones,
and thus we have a strong indication that this correspondence is not an
artifact of the spacetime symmetries but rather it arises from the features of
the two theories, a general proof (or disproof) in the case of arbitrary
metrics is still needed. In order to handle this issue, one might use the
known relation  between Ho\v{r}ava-Lifshitz gravity and Einstein-aether theory
\cite{Jacobson:2010mx,Donnelly:2011df,Yagi:2013ava}.

It is interesting to mention that Gravity's Rainbow, in the FLRW background,
generates Ho\v{r}ava-Lifshitz gravity under a specific form of $f\left(
R\right)  $ theory, with $R$ the 3-dimensional scalar curvature. A similar
result was pointed out in \cite{Olmo:2011sw}, where a connection between the
rainbow's functions and a specific $f\left(  R\right)  $ form seems to be
evident. In our analysis we saw that the obtained correspondence includes
information even for the terms of the type $R^{ij}R_{ij}$, $RR^{ij}R_{ij}$ and
$R_{j}^{i}R_{k}^{j}R_{i}^{k}$ that were not explicitly included.
Hence, we deduce that in order to incorporate higher curvature terms, it is
likely that the rainbow's functions must include terms of the form
$R^{ij}R_{ij}$ etc, a possibility that could be encoded in the Kretschmann
scalar. 
These issues reveal that bridge between Gravity's Rainbow and
Ho\v{r}ava-Lifshitz gravity could be much richer, and deserves further investigation.

We close this work by  mentioning that in the above analysis we have remained at the 
background level, as a first step towards bridging the two theories. However, it is both required and
interesting to examine their relation at the perturbation level too, since there are many example of theories that 
coincide at the background level, while being distinguishable or different when one incorporates the perturbations.
Furthermore, relating the perturbations between Gravity's Rainbow and
Ho\v{r}ava-Lifshitz gravity becomes necessary having in mind the problems of the extra mode propagation that
appears in the simple versions of the latter \cite{Charmousis:2009tc,Li:2009bg,Bogdanos:2009uj,Koyama:2009hc}.
Since such a detailed analysis lies beyond the scope  of the present manuscript
it is left for a future investigation.

 \begin{appendix}

\section{Kinetic term in Gravity's Rainbow with a time-dependent energy term}

\label{App} In the case where $E\equiv E\left(  a\left(  t\right)  \right)  $
the extrinsic curvature of the metric (\ref{FLRWMod}) acquires the form of
relation (\ref{ExrCurvGrw}), namely
\begin{align}
& K_{ij}=-\frac{g_{1}\left(  E\left(  a\left(  t\right)  \right)
/E_{P}\right)  }{2N}\frac{d}{dt}\left[  \frac{g_{ij}}{g_{2}^{2}\left(
E\left(  a\left(  t\right)  \right)  /E_{P}\right)  }\right]  \nonumber\\
& \ \ \ \ \ \,=\frac{g_{1}\left(  E\left(  a\left(  t\right)  \right)
/E_{P}\right)  }{g_{2}^{2}\left(  E\left(  a\left(  t\right)  \right)
/E_{P}\right)  }\left[  \tilde{K}_{ij}+\tilde{g}_{ij}\frac{A\left(  t\right)
}{N\left(  t\right)  }\dot{a}\left(  t\right)  \right]  ~,
\end{align}
where%
\begin{equation}
A\left(  t\right)  =\frac{1}{g_{2}\left(  E\left(  a\left(  t\right)  \right)
/E_{P}\right)  E_{P}}\frac{d}{dE}\Big[g_{2}\left(  E\left(  a\left(  t\right)
\right)  /E_{P}\right)  \Big]\frac{dE}{da},
\end{equation}
and with dots denoting differentiation with respect to time. In the above
expressions   the tildes indicate the quantities computed in absence
of the rainbow's functions.  The trace of the extrinsic curvature
becomes
\begin{equation}
K=g^{ij}K_{ij}=g_{2}^{2}\left(  E\left(  a\left(  t\right)  \right)
/E_{P}\right)  \tilde{g}^{ij}K_{ij}=g_{1}\left(  E\left(  a\left(  t\right)
\right)  /E_{P}\right)  \left[  \tilde{K}+3\frac{A\left(  t\right)  }{N\left(
t\right)  }\dot{a}\left(  t\right)  \right]  ,
\end{equation}
while raising the indices in $K_{ij}$ we obtain
\begin{equation}
K^{ij}=g^{il}g^{jm}K_{lm}=g_{2}^{2}\left(  E\left(  a\left(  t\right)
\right)  /E_{P}\right)  g_{1}\left(  E\left(  a\left(  t\right)  \right)
/E_{P}\right)  \left[  \tilde{K}^{ij}+\tilde{g}^{ij}\frac{A\left(  t\right)
}{N\left(  t\right)  }\dot{a}\left(  t\right)  \right]  .
\end{equation}
Hence, the kinetic term becomes
\begin{align}
& \!\!\!\!\!\!\!\!\!\!\!\!\!\!\!\!\!\!\!\!\!\!\!\!\!\!\!K^{ij}K_{ij}-\lambda
K^{2}=g_{1}^{2}\left(  E\left(  t\right)  /E_{P}\right)  \left\{  \tilde
{K}^{ij}\tilde{K}_{ij}-\lambda\tilde{K}^{2}\right.  \nonumber\\
& \ \ \ \ \ \ \ \ \ \ \ \ \ \ \ \ \ \ \ \ \ \ \ \ \ \ \ \ \  \left.  +\left(
1-3\lambda\right)  \left\{  \frac{2\tilde{K}}{N\left(  t\right)  }A\left(
t\right)  \dot{a}\left(  t\right)  +3\left[  \frac{A\left(  t\right)
}{N\left(  t\right)  }\dot{a}\left(  t\right)  \right]  ^{2}\right\}
\right\}  .\label{Kinetic}%
\end{align}
For the specific case of a FLRW metric we find that
\begin{equation}
\tilde{K}_{ij}=-\frac{\tilde{g}_{ij}}{N\left(  t\right)  }\frac{\dot{a}}{a},
\end{equation}
and thus
\begin{equation}
\tilde{K}^{ij}\tilde{K}_{ij}-\lambda\tilde{K}^{2}=3\frac{\left(
1-3\lambda\right)  }{N^{2}\left(  t\right)  }\left(  \frac{\dot{a}}{a}\right)
^{2}.
\end{equation}
In this case Eq. $\left(  \ref{Kinetic}\right)  $ becomes
\begin{equation}
K^{ij}K_{ij}-\lambda K^{2}=3g_{1}^{2}\left(  E\left(  t\right)  /E_{P}\right)
\frac{\left(  1-3\lambda\right)  }{N^{2}\left(  t\right)  }\left(  \frac
{\dot{a}}{a}\right)  ^{2}f\!\left(  A\left(  t\right)  ,a\right)
,\label{Kin1}%
\end{equation}
where
\begin{equation}
f\!\left(  A\left(  t\right)  ,a\right)  =\left[  1-2a\left(  t\right)
A\left(  t\right)  +A^{2}\left(  t\right)  a\left(  t\right)  ^{2}\right]
.\label{f(A(t),a)}%
\end{equation}
It is now possible to calculate the kinetic part of the action, which is
defined as%
\begin{equation}
S_{K}=\int_{\Sigma\times I}dtd^{3}x\mathcal{L}_{K},
\end{equation}
where
\begin{equation}
\mathcal{L}_{K}=\frac{N}{2\kappa}\sqrt{g}\left(  K^{ij}K_{ij}-\lambda
K^{2}\right)  .
\end{equation}
Inserting $\left(  \ref{Kin1}\right)  $ into $S_{K}$ we obtain
\begin{equation}
S_{K}=\frac{3}{\kappa}\pi^{2}\int_{I}dtN\left(  t\right)  a\dot{a}^{2}%
\,\frac{g_{1}^{2}\left(  E\left(  a\left(  t\right)  \right)  /E_{P}\right)
}{g_{2}^{3}\left(  E\left(  a\left(  t\right)  \right)  /E_{P}\right)  }%
\frac{\left(  1-3\lambda\right)  }{N^{2}\left(  t\right)  }f\!\left(  A\left(
t\right)  ,a\right)  ,
\end{equation}
and thus the canonical momentum reads as
\begin{equation}
\pi_{a}=\frac{\delta S_{K}}{\delta\dot{a}}=\frac{g_{1}^{2}\left(  E\left(
a\left(  t\right)  \right)  /E_{P}\right)  }{g_{2}^{3}\left(  E\left(
a\left(  t\right)  \right)  /E_{P}\right)  }f\!\left(  A\left(  t\right)
,a\right)  \tilde{\pi}_{a},\label{Pia}%
\end{equation}
where%
\begin{equation}
\tilde{\pi}_{a}=\frac{6\pi^{2}}{\kappa}\frac{\left(  1-3\lambda\right)
}{N\left(  t\right)  }\dot{a}a.
\end{equation}
Definitely, we restrict ourselves in the case $\lambda\neq\frac{1}{3}$, since
in the special case where $\lambda=\frac{1}{3}$ the ultralocal metric (the one-parameter family of
supermetrics which allows to disentangle gauge modes from physical
deformations) \cite{MM,Giulini} is not invertible and becomes a projector onto
the tracefree subspace.

\section{The Lichnerowicz equation for the graviton}

\label{Appen} In $3+1$ dimensions the graviton operator is described by
\begin{equation}
O^{ikjl}=\bigtriangleup_{L}^{ikjl}-4R^{il}g^{kj}+Rg^{ik}g^{jl}+\frac
{\partial^{2}}{\partial t^{2}}g^{ik}g^{jl},
\end{equation}
where we have assumed the absence of mixing between time and space, which
naturally follows from the structure of the FLRW metric (\ref{FLRW}). The
Riemann tensor in $3$ dimensions becomes%
\begin{equation}
R_{ikjl}=g_{ij}R_{kl}-g_{il}R_{kj}-g_{kj}R_{il}+g_{kl}R_{ij}-\frac{R}%
{2}\left(  g_{ij}g_{kl}-g_{il}g_{kj}\right)  ,
\end{equation}
and for a FLRW background the 3-dimensional Ricci curvature tensor and the
scalar curvature read%
\begin{equation}
R_{ij}=\frac{2}{a^{2}\left(  t\right)  }\gamma_{ij}\qquad\mathrm{and}\qquad
R=\frac{6}{a^{2}\left(  t\right)  }~,
\end{equation}
where $\gamma_{ij}$ is the metric on the spatial sections which have constant
curvature $k=0,\pm1$, defined by
\begin{equation}
d\Omega_{3}^{2}=\gamma_{ij}dx^{i}dx^{j}~.
\end{equation}
Hence, the Riemann tensor reduces to%
\begin{equation}
R_{ikjl}=-\frac{2}{a^{2}\left(  t\right)  }\left(  \gamma_{ij}\gamma
_{kl}-\gamma_{il}\gamma_{kj}\right)  .
\end{equation}
Then, the operator $O^{ikjl}$ on  transverse traceless tensors  reduces
to%
\begin{equation}
O^{ikjl}=a^{-2}\left(  t\right)  \left(  -\nabla^{a}\nabla_{a}\gamma
^{ik}\gamma^{jl}+2\gamma^{il}\gamma^{kj}\right)  +\frac{1}{N^{2}}%
\frac{\partial^{2}}{\partial t^{2}}\gamma^{ik}\gamma^{jl},
\end{equation}
and the dispersion relation becomes%
\begin{equation}
\frac{k^{2}}{a^{2}\left(  t\right)  }=E^{2},\label{p31}%
\end{equation}
where as usual in the end of the calculation we have set the lapse function
$N$ to $1$. Finally, as it was shown in \cite{GaMa}, in the case of Gravity's
Rainbow the above dispersion relation has to be modified to
\begin{equation}
\frac{k^{2}}{a^{2}\left(  t\right)  }=\frac{E^{2}}{g_{2}^{2}\left(
E/E_{P}\right)  }.\label{MDR}%
\end{equation}
 \end{appendix}

\end{document}